\documentstyle[12pt]{article}
\begin{document}
\begin{titlepage}
 
\vspace{2.5cm}
\begin{centering}
{\Large{\bf Electrostatic self-force in (2+1)-dimensional cosmological gravity}}\\
\bigskip\bigskip
Cl\'audio Furtado and Fernando Moraes\\
{\em Departamento de F\'{\i}sica\\
Universidade Federal de Pernambuco\\
50670-901 Recife, PE, Brazil}\\
\vspace{1cm}
\end{centering}
\vspace{1.5cm}

\begin{abstract}

Point sources in (2+1)-dimensional gravity are conical singularities that modify the global curvature of the space giving rise to self-interaction effects on classical fields. In this work we study the electrostatic self-interaction of a point charge in the presence of point masses in (2+1)-dimensional gravity with a cosmological constant $\Lambda$. Exact expressions are obtained for the self-force on the charge in both $\Lambda < 0$ and $\Lambda > 0$ cases, which correspond to locally hyperbolic and spherical spaces, respectivelly. Both results, in the limit $\Lambda \rightarrow 0$, coincide with the well-known 
flat space case.
\end{abstract}

\hspace{0.8 cm}
\end{titlepage}
\def\carre{\vbox{\hrule\hbox{\vrule\kern 3pt
\vbox{\kern 3pt\kern 3pt}\kern 3pt\vrule}\hrule}}

\baselineskip = 18pt
\section{Introduction}
Lower dimensional gravity besides its pedagogical role~\cite{Jac} has important applications in condensed matter: elastic solids with defects can be mapped into three-dimensional gravity with torsion~\cite{Kat}. In this framework, we have studied ~\cite{Fur,Mor} a few cases concerning classical and quantum effects in solid media containing topological defects such as disclinations and dislocations (a disclination in a solid is analogous to a cosmic string). An important issue in this class of problems is the question of the self-interaction of a point charge in the background space of a topological defect. This is a particular case of the well known~\cite{DeW} fact that a point charge in a static gravitational field experiences an electrostatic force due to the change in the boundary conditions its electric field is submited to. In other words, the change in the geometry of the space-time which is associated with the gravitational field causes a deformation of the electromagnetic field lines  inducing a self

-force on the charge. Linet~\cite{Lin} and Smith~\cite{Smi} have found the electrical self-force on a point charge in the presence of a cosmic string. de Mello et al.~\cite{Eug} found the magnetic self-force on a electric current parallel to a cosmic string. Souradeep and Sahni~\cite{Sou} and Guimar\~aes and Linet~\cite{Mex} studied the electric self-interaction in a three-dimensional flat background with a conical defect, which is equivalent to 2+1 gravity without a cosmological constant $\Lambda$~\cite{Jac}. This is so~\cite{Des} because the gravitational field equations in 2+1 dimensions with a cosmological constant give a space-time of constant curvature as solution. Depending on the sign of $\Lambda$ the result is a closed $R^1 \times S^2$ de Sitter space ($\Lambda > 0$) or a  $R^1 \times H^2$ hyperbolic anti-de Sitter space. In fact, the radius of curvature of the space is
\begin{equation}
R=\frac{1}{\sqrt \Lambda}.
\end{equation}
The imaginary radius corresponding to the hyperbolic case.

In this work we consider the self-force effect on a point charge in the presence of a point mass in a (2+1)-dimensional spacetime with cosmological constant; i.e. we find the self-force on an electric charge in $R^1 \times S^2$ and in $R^1 \times H^2$ with  conical singularities. The introduction of a point source of gravitational field in a (2+1)-dimensional space-time is equivalent to the creation of a conical singularity in the otherwise smooth background.  The conical singularity appears as a result of Voterra's process~\cite{Kle1} of cutting out a sector of the space and identifying the loose ends.  In flat space this results in a single source but in  the sphere its topology generates two point masses~\cite{Des} antipodal to each other. By choosing the coordinate system such that the defects coincide with the ``north" and ``south" poles, the inclusion of the pair of point masses is equivalent to changing the periodicitity of the azimuthal angle $\phi$ from $2\pi$ to $2\pi /p$, where~\cite{Sou}
\begin{equation}
p=(1-4G_{2}M)^{-1}, (p\geq 1).
\end{equation}
Here, $M$ is the mass of the point source and $G_2$ is the gravitational constant in 2+1 dimensions. The same is also true for the flat and hyperbolic cases, except that there is only one point mass and it is located at the origin.

We basically follow the procedure used by Smith~\cite{Smi} in order to compute the self-forces. It goes as follows: we first solve Poisson's equation for a single charge in the chosen background in order to find its Green function. The Green function is then renormalized by the extraction of its divergent part, namely the Green function in the absence of the point mass. The self-energy of the point charge will be then~\cite{Smi}
\begin{equation}
U(x)=\frac{1}{2}q^{2}G_{p}(x,x)_{ren}
\end{equation}
where $x$ represents the two-dimensional coordinates of a point in the space section of our manifold and $q$ is the charge of the particle.

\section{Spherical space ($\Lambda > 0$)}
Like the mass source, a single electric charge is not allowed in spherical space. Since the field lines will be closed due to the compactness of the space, an antipodal equal charge of opposite sign must also be present. The interaction between them is obtained from Eq. (16) below, by considering their mutual potential energy to be $\frac{1}{2}G_{p}(x,x')$. Since we are interested in the gravitational effects of the pair of point masses on the electrostatics in a spherical space, we focus on the self-interaction of one of the charges only. Notice that in the flat background space limit ($\Lambda \rightarrow 0$), if we are to keep one of the charges and one of the point masses, their respective antipodes will be at infinity.
 
The metric of $S^2$ is given by
\begin{equation}
dS^{2} = R^{2} d \theta^{2} + R^{2} \sin^{2}\theta d \phi^{2},
\end{equation}
where $R=\frac{1}{\sqrt \Lambda}$ is the radius of the sphere. The inclusion of the point masses makes $0\leq \phi \leq 2\pi /p$ but does not change the range of $\theta$, which stays $0 \leq \theta \leq \pi$. 

We need to solve Poisson's equation 
\begin{equation}
\Delta G = - 2 \pi \delta (x - x'),
\end{equation}
where the Laplacian $\Delta$ is given by
\begin{equation}
\Delta = \frac{1}{R^{2} \sin \theta} \frac{\partial}{\partial \theta}
\sin \theta \frac{\partial}{\partial \theta} +
\frac{1}{R^{2}\sin^{2} \theta} \frac{\partial^{2}}{\partial \phi^{2}}
\end{equation}
and
\begin{eqnarray*}
\delta (x - x') = \frac{1}{R^{2} \sin \theta} \delta
( \theta - \theta ' ) \delta ( \phi - \phi ').
\end{eqnarray*}
Expanding $\delta(\phi-\phi ')$ in Fourier components, we get
\begin{equation}
\delta (\phi - \phi ' ) = \frac{p}{\pi} \sum^{\infty}_{m=0} a_{m}
\cos [mp ( \phi - \phi ')],
\end{equation}
where
\begin{eqnarray*}
a_{m} = \left \{ \begin{array}{l}
\frac{1}{2} \; \mbox{if} \; m=0 \\
1 \; \mbox{if} \; m>0 \end{array} \right.
\end{eqnarray*}

Now, using the {\it ansatz} 
\begin{equation}
G({x},{x}') = \frac{p}{\pi}
\sum^{\infty}_{m=0} a_{m} \cos [pm ( \phi - \phi ')] g_{m} (
\theta , \theta ' )
\end{equation}
in Poisson's equation, we find
\begin{equation}
\frac{1}{R^{2}\sin \theta } \frac{\partial}{\partial \theta} \sinh
\theta \frac{\partial}{\partial \theta} g_{m} ( \theta , \theta
' ) - \frac{p^{2}m^{2}}{R^{2}\sin^{2} \theta} g_{m} ( \theta ,
\theta ' ) = - \frac{2 \pi \delta (\theta - \theta ')}{R^{2} \sin
\theta}.
\end{equation}

For $m\neq 0$ Equation (9) gives
\begin{equation}
g_{m} ( \theta , \theta ') = \frac{\pi}{mp}
\left( \frac{\cot ( \theta_{>}/2}{\cot (\theta_{<}/2)} \right)^{mp}, 
\end{equation}
where $\theta_{<}$ ($\theta_{>}$) is the smaller (larger) of $\theta$ and $\theta'$. For $m=0$ the solution is
\begin{equation}
g_{0} ( \theta , \theta ' ) = - 2 \pi \ln | \tan \theta /2 |.
\end{equation}
Now, substituting $g_m$ in Eq. (8) results in
\begin{eqnarray}
G(x,x') & = & \sum^{\infty}_{m=0} 
\frac{\cos [mp ( \theta - \theta ')]}{m} \; 
\left[ \frac{\cot (\theta_{>}/2)}{\cot (\theta_{<}/2)} \right]^{mp}\nonumber\\[0.3cm]
& - & p \ln | \tan \theta_{>}/2 |.
\end{eqnarray}
In order to perform the sum in the above equation we make
\begin{equation}
X = \frac{\cot ( \theta_{>}/2}{\cot (\theta_{<}/2)}. 
\end{equation}
This leads to 
\begin{eqnarray}
G(x,x') & = & \sum^{\infty}_{m=1} \frac{\cos [mp ( \phi -
\phi ')]}{m} X^{m}\nonumber\\[0.3cm]
& - & p \ln | \tan \theta_{>}/2 |.
\end{eqnarray}
Now, using~\cite{Gra}
\begin{eqnarray}
& & \sum^{\infty}_{m=1} \frac{\cos [pm(\phi - \phi ')]}{m} X^{mp}
= \nonumber\\[0.3cm]
& & - \frac{1}{2} \ln \left\{ 1 + X^{2p} - 2X^{p} \cos p ( \phi
- \phi ') \right\},
\end{eqnarray}
we finally get
\begin{eqnarray}
G_{p}(x,x') = - \frac{1}{2} \ln\{ \tan^{2p} (\theta /2) + \tan^{2p} 
(\theta '/2)\nonumber\\[0.3cm] 
-2 \tan^{p} (\theta /2) \tan^{p} ( \theta'/2) \cos p(\phi - \phi ') \}. 
\end{eqnarray}

In order to compute the self-energy of the point charge, Eq. (3), we need to renormalize $G_{p}(x,x')$:
\begin{equation}
G_{p}(x,x')_{ren} = \lim_{x' \rightarrow x} \left\{ G_{p}
(x,x') - G_{1}(x,x') \right\}=  - \frac{1}{2} \ln \left\{ p^{2} \tan^{2p-2}
\theta /2 \right\}.
\end{equation}
The self-energy is then
\begin{equation}
U(x)=- \frac{q^{2}}{2} \ln \left\{ p^{2} \tan^{2p-2}
\theta /2 \right\}.
\end{equation}
The resulting self-force is therefore
\begin{equation}
F = - \nabla U = \frac{(p-1)}{R} q^{2} \frac{1}{\sin \theta}\hat{\theta}.
\end{equation}

\section{Hyperbolic space ($\Lambda < 0$)}
Here the space is $H^2$ with metric given by
\begin{equation}
dS^{2} = d \rho^{2} + R^{2} \sinh^{2} \rho/R d \phi^{2}.
\end{equation}
Again,  $0\leq \phi \leq 2\pi /p$. Here $0 \leq \rho \leq \infty$. We essentially follow the steps of Section 2: the equation corresponding to Eq. (8) is here
\begin{equation}
G_{p}(x,x') = \frac{p}{\pi} \sum^{\infty}_{m=0} a_{m} \cos [mp (
\phi - \phi )] g_{m} ( \rho , \rho ').
\end{equation}
We obtain the following equation for $g_{m}(\rho,\rho ')$:
\begin{eqnarray}
\frac{1}{\sinh \rho /R} \frac{\partial}{\partial \rho} \sinh \rho
/R \frac{\partial}{\partial \rho} g_{m} ( \rho , \rho ') -
\frac{p^{2}m^{2}}{R^{2} \sinh^{2} \rho /R} g_{m} ( \rho , \rho ')\nonumber\\[0.3cm]
= - \frac{2 \pi \delta ( \rho - \rho ')}{R \sinh \rho /R},
\end{eqnarray}
whose solution is
\begin{equation}
g_{m} ( \rho , \rho ') = \frac{\pi}{mp} \left( \frac{\coth (
\rho_{>} /2R )}{\coth ( \rho_{<} / 2R )} \right)^{mp},
\end{equation}
for $m \neq 0$ and
\begin{equation}
 g_{0} ( \rho , \rho ') = -2 \pi \ln | \tanh \rho / 2R |,
\end{equation}
for $m=0$.

Substituting Eqs. (23) and (24) in Eq. (21) and performing the sum, leads to
\begin{eqnarray}
G_{p}(x,x') = & -&\frac{1}{2} \ln\{ \tanh^{2p} ( \rho / 2R) + \tanh^{2p} (\rho ' /2R) \nonumber\\[0.3cm] 
&-&2\tanh^{p} (\rho /2R) \tanh^{p} (\rho '/2R) \cos p ( \phi - \phi ')\}, 
\end{eqnarray}
which in its renormalized form is
\begin{equation}
G_{p}(x,x')_{ren} = - \frac{1}{2} \ln \left\{ p^{2} \tanh^{2p-2} (\rho
/2R) \right\}. 
\end{equation}
The self-energy and self-force are respectively
\begin{equation}
U = - \frac{q^{2}}{2} \ln \left\{ p^{2} \tanh^{2p-2} (\rho /2R)
\right\}
\end{equation}
and
\begin{equation}
F = (p-1) \frac{q^{2}}{R} \frac{1}{\sinh (\rho /R)} \hat{\rho}.
\end{equation}

\section{The method of images}
In this section we rederive Eq. (25) in the more intuitive approach of the method of images~\cite{Smi,Sou} for the special case where $p$ is a positive integer. Although we do the calculation explicitly for the hyperbolic case the spherical case is analogous.

The point charge will have $p-1$ images distributed around the origin at $\phi = \frac{2\pi k}{p}$ for $k= 1, 2, ..., p-1$. The renormalized Green function is obtained from~\cite{Sou}
\begin{eqnarray}
G_{p}(\rho,\phi,\rho ',\phi ')_{ren}= \sum^{p-1}_{k=1}G_{1}(\rho,\phi,\rho ',\phi '+\frac{2\pi k}{p})=-\frac{1}{2} \sum^{p-1}_{k=1} \ln\{ \tanh^{2}( \rho / 2R)\nonumber\\[0.3cm]
+\tanh^{2} ( \rho ' /2R) - 2 \tanh \rho /2R \tanh \rho ' /2R \cos ( \phi - \phi ' +\frac {2\pi k}{p})\}. 
\end{eqnarray} 
Using~\cite{Gra}
\begin{equation}
\sum^{+\infty}_{m=-\infty}\,'\,\,\frac{X^{|m|}}{m}e^{im(\phi - \phi ')}=\ln [1+X-2X\cos (\phi - \phi ')],
\end{equation}
where
\begin{equation}
X=\frac{\tanh (\rho '/2)}{\tanh (\rho /2)},
\end{equation}
and $\sum\,'$  indicates that the value $m=0$ is not included in the sum, we easily obtain Eq. (25).
 
\section{Conclusion}
We have obtained the self-interaction expressions for a point charge in the gravitational background of point sources in a (2+1)-dimensional universe with cosmological constant $\Lambda \neq 0$. Topological restrictions force the $\Lambda > 0$ case ($S^2$) to have a minimum of two antipodal sources. In the $\Lambda < 0$ case ($H^2$) a single point source is allowed by topology (as in the $\Lambda = 0$ case). Although we performed the calculations separately for each case, they can be mapped into each other by changing the sign of $\Lambda$ or, equivalently, taking the radius of curvature $R$ into $iR$ and defining $\rho = R \theta$. The Euclidean background ($\Lambda \rightarrow 0$) limit~\cite{Eug,Sou,Mex}
\begin{equation}
F = (p-1) \frac{q^{2}}{r} \hat{r},
\end{equation}
is easily obtained from both the spherical and hyperbolical cases by making $\theta = r/R$ and $\rho = r$, respectively. Both in the hyperbolic mapping or in the Euclidean limit of the spherical case one of the two sources is preserved while its antipodal is moved to infinity.

A final comment concerning condensed matter applications: amorphous solids have been described~\cite{Kle2} by regular arrangements of atoms in a constant curvature space with disclinations. The results of this work may therefore be of use to the study of electronic properties of amorphous solids along the lines of Ref.~\cite{Fur}.

\noindent
{\bf Acknowledgment}\\
\noindent
This work was partially supported by CNPq.

\end{document}